\documentclass{article}
\usepackage{graphicx}
\usepackage{latexsym}

\begin{document}

\begin{flushright}
hep-ph/0304216\\
KIAS-P03031\\
UM-P006-2003
\end{flushright}

\begin{center}
{\bf {\huge Neutrino lensing and modification of Newtonian gravity
at large distances.}}\\
\hspace{10pt}\\
S. R. Choudhury\footnote{src@physics.du.ac.in} \\
{\em Department of Physics, Delhi University, Delhi 110007, India},\\
\hspace{10pt}\\
A. S. Cornell\footnote{a.cornell@tauon.ph.unimelb.edu.au}\\
{\em Korea Institute of Advanced Study, 207-43 Cheongryangri
2-dong, Dongdaemun-gu, Seoul 130-722, Republic of Korea},\\
\hspace{10pt}\\
and G. C. Joshi\footnote{joshi@physics.unimelb.edu.au}\\
{\em School of Physics, University of Melbourne,}\\
{\em Victoria 3010, Australia}\\
\hspace{10pt}\\
$23^{rd}$ of April, 2003
\end{center}
\hspace{10pt}\\

\begin{abstract}
\indent The nature of the modification to neutrino lensing from
galaxies, as caused by possible modifications to Newtonian gravity
at large distances, is studied.
\end{abstract}

\section{Introduction}
\hspace{10pt} The validity of Newtonian gravity laws have been
experimentally verified to distances as short as about a tenth of
a millimeter \cite{Hoyle2001}.  On the other hand there has been
no direct experimental verification beyond stellar scale
distances.  Possible modification of Newtonian gravity both at
small and large distances has become the subject of many new
theories \cite{Arkani1998}.  These are theories in which the
universe is no longer four dimensional, the extra dimensions being
invoked to solve the so called `hierarchy problem' between scales
involved in the Standard Model of particle interactions and the
four dimensional `Planck Scale' \cite{Randall1999}.  Kogan et al.
\cite{Kogan2001} proposed in this connection a three brane model
of the universe which removes some of the drawbacks of the earlier
brane model of Randall and Sundrum \cite{Randall1999}, which had
only two branes.  This model has the distinct feature, unlike the
two brane model of Randall and Sundrum, of having possible
modifications to Newtonian gravity at large distances into a
Yukawa type of interaction, rather than the $1/r$ gravitational
potential.  The theory of course gives no indication of the range
of such a Yukawa type potential.  The Yukawa type of gravitational
interaction also results from other approaches, like one involving
higher terms in the gravitational interaction \cite{Steele1978}.
Phenomenologically the modification to Newtonian gravity over
large distances can be studied through the lensing of light and
neutrinos, where we have their origins as far away from us.
Lensing of optical images has of course been studied widely in the
context of standard Einsteinian gravity.  Some time back McKellar,
Mahajan and two of us (SRC, GCJ)\cite{Choudhury2002}, studied the
possibility of using detailed data from the lensing of distant
stars as a tool for studying possible modifications of Newtonian
gravity at cosmological scales.  Lately, neutrino lensing and
their possible observation has attracted much attention.  In two
recent papers Escribano, Frere, Monderen and Van Elewyck
\cite{Escribano2002} have performed detailed analysis of
magnitudes of neutrino lensing that can be expected from normal
gravitational interactions with sources such as stars and
galaxies.  In line with our earlier study we devote this note to
explicitly recording the nature of the modification of such
lensing predictions of Newtonian gravity for a sphere of radius R
with a finite range Yukawa type interaction with a range in the
cosmological scale.\\
\indent The energy-momentum tensor of a spherically symmetric mass
of radius R is given by,
\begin{equation}
T_{\mu \nu} = p  g_{\mu \nu} + ( p + \rho ) U_{\mu} U_{\nu}
\end{equation}
where $U_{\mu}$ is the four velocity.  Use of such an
energy-momentum tensor in the standard Einstein equations leads to
a Newtonian gravity at large distances from the sphere.  We use an
ad hoc prescription  to simulate an effective Yukawa type of
gravity rather than Newtonian by adding to $T_{\mu \nu}$ above an
extra term such that the new energy momentum reads,
\begin{equation}
T^{\prime}_{\mu \nu } = T_{ \mu \nu } + \frac{\Lambda(r)}{r^2}
g_{\mu\nu}
\end{equation}
and suitably adjust the value of $\Lambda(r)$ so that at large
distances we end up with Yukawa type of gravity.  Writing the four
dimensional line element in the standard form\\
\begin{equation}
ds^2 = B(r) dt^2 - A(r) dr^2 - r^2 d\theta^2  - r^2 \sin^2\theta
d\phi^2
\end{equation}
we obtain the following equations for the components of the
curvature tensor\\
\begin{eqnarray}
R_{rr} &=& - 4 \pi G ( \rho - p ) A (r) +  8 \pi G
\frac{\Lambda(r)}{r^2} ,
\\
R_{\theta\theta} &=& - 4 \pi G ( \rho - p ) r^2 + 8 \pi G \Lambda(r) , \\
R_{tt} &=& - 4 \pi G ( \rho + 3 p) B(r)  - 8 \pi G
\frac{\Lambda(r)}{r^2} .
\end{eqnarray}
In the matter free region, $\rho = p = 0$, we continue to have
\begin{equation}
\frac{R_{rr}}{A(r)}  + \frac{ R_{tt}}{B(r)} = 0
\end{equation}
so that in this region we still have $A(r) = 1/B(r)$.\\
\indent In the presence of matter the equation for $A(r)$, by
standard manipulation of equations (4)-(6) to eliminate pressure,
becomes,
\begin{equation}
\frac{d}{dr}\left(\frac{r}{A(r)}\right)= 1 - 8 \pi G \rho r^2 + 8
\pi G \Lambda(r) .
\end{equation}
We can now assume a form for $\Lambda (r)$;
\begin{equation}
\Lambda(r) =  \Lambda e^{-\lambda r}
\end{equation}
with $\lambda$ and $\Lambda$ being constants.  Equation (8) is
then easily integrated yielding
\begin{equation}
\frac{ 1} {A(r)} = 1 - \frac{ 2 G M(r)}{r} - \frac{ 8 \pi G
\Lambda}{\lambda r } \left(e^{-\lambda r} -1 \right)
\end{equation}
where the constant of integration has been adjusted to make $A(r)$
finite at r=0 and $M(r)$ is the mass contained between $r=0$ and
$r$.  Identifying
\begin{equation}
8 \pi \Lambda = 2 M \lambda
\end{equation}
we get for values of $r$ outside $R$
\begin{equation}
\frac{1}{A(r)} = 1 - \frac{ 2 G M }{r}  e^{-\lambda r}
\end{equation}
which is the Yukawa form of gravitation that we desired to
reproduce.  The form of $A(r)$ given in equation (10) is of course
valid for all $r$ and for values of $r$ much smaller than $R$, it
can be approximated by
\begin{equation}
\frac{1}{A(r)} = 1 - \frac{2 G M(r)}{r} + 2 G \lambda .
\end{equation}
The value of the function $B(r)$ for $r > R$ is of course the
inverse of $A(r)$.  To obtain values of the function $B(r)$ inside
the spherical core, we assume that the matter distribution in the
core is Gaussian, in which case we can make the approximation
$\rho << p$ and the function $B(r)$ there satisfies
\begin{equation}
\frac{ B^\prime}{B} = \frac{ 2 M (r)}{ r^2} .
\end{equation}
As we have assumed a Gaussian matter distribution the density
inside the core is of the form
\begin{equation}
\rho(r) = \rho_o e^{- r^2/r_o^2} .
\end{equation}
The mass $M(r)$ for such a density turns out to be
\begin{eqnarray}
M(r) &=& \frac{M}{R}  r  e^{\frac{(R^2 - r^2)}{r_o^2}} \frac{f(r_o/r)}{f(r_o/R)}   \\
f(x) &=& x e^{x^2} \frac{\sqrt{\pi}}{2} \mathrm{erf}(x) -1
\nonumber .
\end{eqnarray}
For $r < R$ the function $B(r)$ is then
\begin{equation}
B(r) = 1 - \frac{2 M}{R} e^{\frac{(R^2 - r^2)}{r_o^2}}
\frac{f(r_o/r)}{f(r_o/R)} + O (M^2)
\end{equation}
\indent Given the functions $A(r)$ and $B(r)$, the deflection of a
beam of neutrinos (assumed massless) initially moving along the
direction $\phi=0$ is given by
\begin{equation}
\phi(r) = \displaystyle \int^r_\infty \frac{A^{1/2}(r) dr}{r^2}
\left(\frac{1}{b^2 B(r)} - \frac{1}{b^2} - \frac{1}{r^2} \right)
\end{equation}
where $b$ in the equation above is the impact parameter.  The total
deviation of the beam is then given by\\
\begin{equation}
\Delta\phi = 2 \phi ( r=r_{min} ) - \pi .
\end{equation}
For neutrino trajectories which lie totally outside the core, the
above equation simplifies since $A(r) = 1/B(r)$ there.  We get
\begin{eqnarray}
y&=&\frac{1}{r} - M \frac{e^{-\lambda/r}}{r^2} \\
\phi(y)&=& \displaystyle \int^y_0 dy \frac{1 + (2 y - \lambda) M
e^{-\lambda/y}}{\sqrt{1/b^2  - y^2 }} +O(M^2) .
\end{eqnarray}
The maximum value of $y$ is $1/b$ and the integral above can be
easily numerically integrated to give the value of $\Delta\phi$ by
equation (19).  For trajectories that pass through the region
$r<R$, one has to numerically integrate using the value of $A(r)$
from equation (13) above and $B(r)$ from equation (17).\\
\indent We display our results in figure 1 for a typical galactic
mass and radius.  As expected, for neutrino trajectories that come
deep through the mass of the galaxy, the Yukawa nature of the
gravitational force does not cause any change in the deflection of
the beam.  However, there is considerable change in the deflection
of the beam, as expected, if the impact parameter is substantially
greater than the core radius.  Of course, in those regions the
deflection of the beam also decreases, making observations there
more difficult than trajectories through the core.  It seems,
therefore, that it would require some fortuitous circumstances of
a suitable set of sources, from where neutrino beams would be
lensed through galactic masses, for observations to possibly
detect changes in the law of gravitation for neutrinos.  Most
certainly, the optical analog of such observations present a more
promising scenario for future experiments.

\section*{Acknowledgements}
The work of SRC was supported by a project grant from the
Department of Science and Technology, India.  ASC would also like
to thank the department of Physics \& Astrophysics, University of
Delhi, for their support during his visit to India.

\begin{figure}
\begin{center}
\includegraphics[angle=270,width=12cm]{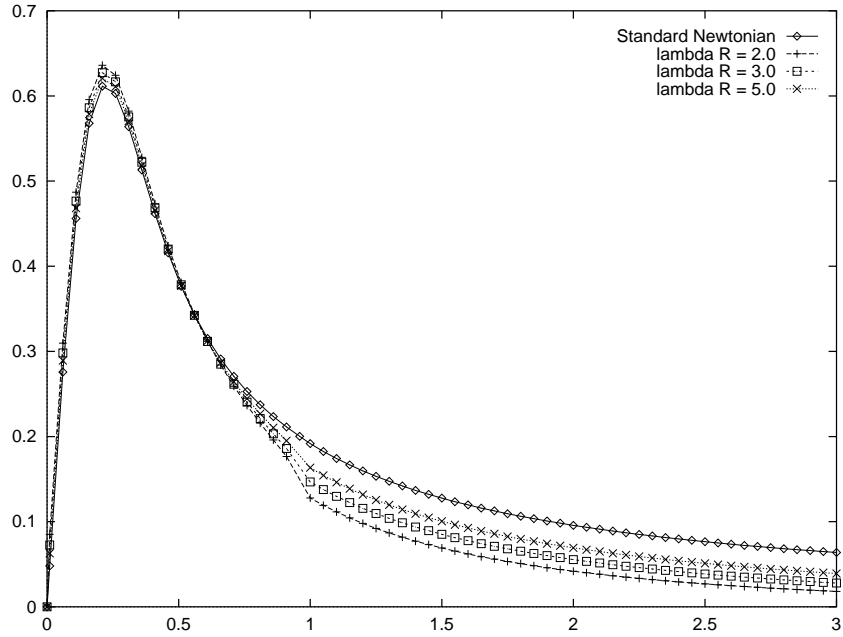} \caption{Graphs showing
the deflection $\Delta\phi$ in units of $10^{-5} radians$ for
various choices of $\lambda R$, where
$M=M_{Galaxy}=10^{12}M_{\odot}$, $R=R_{Galaxy}=100$kpc and
$r_o=0.2 R_{Galaxy}$. Also shown is the graph for standard
Newtonian gravity.} \label{fig2}
\end{center}
\end{figure}

\end{document}